# Guiding Data-Driven Design Ideation by Knowledge Distance

Jianxi Luo[abc*]  Serhad Sarica[a]  Kristin L. Wood[ab]

luo@sutd.edu.sg  serhad_sarica@mymail.sutd.edu.sg  kristinwood@sutd.edu.sg

[a] Singapore University of Technology and Design (SUTD), 8 Somapah Road, 487372, Singapore

[b] SUTD-MIT International Design Centre, 8 Somapah Road, 487372, Singapore

[c] SUTD Design and Artificial Intelligence Programme, 8 Somapah Road, 487372, Singapore

* Corresponding author: Jianxi Luo, luo@sutd.edu.sg, +65 6499 4504

**Acknowledgement**

This research is supported by SUTD-MIT International Design Centre (IDC, idc.sutd.edu.sg, IDG31600105) and Singapore Ministry of Education Tier 2 Academic Research Grant (T2MOE1403).

**Abstract**

Data-driven conceptual design methods and tools aim to inspire human ideation for new design concepts by providing external inspirational stimuli. In prior studies, the stimuli have been limited in terms of coverage, granularity, and retrieval guidance. Here, we present a knowledge-based expert system that provides design stimuli across the semantic, document and field levels simultaneously from all fields of engineering and technology and that follows creativity theories to guide the retrieval and use of stimuli according to the knowledge distance. The system is centered on the use of a network of all technology fields in the patent classification system, to store and organize the world's cumulative data on the technological knowledge, concepts and solutions in the total patent database according to statistically-estimated knowledge distance between technology fields. In turn, knowledge distance guides the network-based exploration and retrieval of inspirational stimuli for inferences across near and far fields to generate new design ideas by analogy and combination. With two case studies, we showcase the effectiveness of using the system to explore and retrieve multilevel inspirational stimuli and generate new design ideas for both problem solving and open-ended innovation. These case studies also demonstrate the computer-aided ideation process, which is data-driven, computationally augmented, theoretically grounded, visually inspiring, and rapid.

# 1. Introduction

Innovators and innovative companies often ask questions such as

- "What new functions and features can be added to our current products to create new products?"
- "What new products should we design next, based on our technologies?"
- "What solutions can we design to solve this problem?"

Novel and feasible answers to these questions may lead to innovation. To seek answers to such innovation questions, structured ideation methods, such as brainstorming, mind mapping TRIZ and design heuristics [1–4] may facilitate creative thinking, although design ideation is also conditioned on the knowledge and experience of the designers. Design thinking, user studies and market research may reveal user needs as design opportunities [5–7], but such processes are often slow and resource consuming.

Recently, *data-driven design* methods have been proposed to provide broad *external stimuli* to inspire *design ideation* [8–13]. External stimuli may inspire humans to think outside of the box to produce novel concepts [14–17] but may also cause bias and fixation [18]. Stimuli need to be carefully chosen for data-driven design, especially when big data sources are used. However, the stimuli exemplified in prior studies have been limited in terms of the stimuli data's granularity (e.g., at the keyword, document, or macro-direction level) and choice and coverage (e.g., a sample of keywords or documents), which are often given arbitrarily without theoretically grounded intelligent recommendation algorithms.

To advance data-driven design inspiration for concept generation, we present an *knowledge-based expert system* that provides design stimuli across the semantic, document and field levels at the same time, selected from the total patent database covering technologies in all known fields and with retrieval recommendation algorithms based on design creativity theories. Two case studies will be presented to illustrate the system by applying it to answer the innovator's questions put forth at the beginning of the paper. The case studies also demonstrate the related *computer-aided ideation* process, which is data-driven, computationally augmented, theoretically grounded, visually inspiring, and rapid. Taken together, the knowledge-based expert system and computer-aided ideation process are aimed at contributing to the growing literature on *data-driven design* and *artificial intelligence for design*, with a focus on technology-based design and creative concept generation at the fuzzy front end of the innovation process.

This paper is organized as follows. In Section 2, we briefly review the relevant literature. Section 3 introduces the knowledge-based expert system in detail. Section 4 presents two case studies followed by a summary discussion in Section 5. Section 6 discusses the limitations and opportunities for system extensions and future research.

## 2. Related Work

Our focus is on data-driven design in the concept generation phase at the fuzzy front end of the innovation process and on providing external data to inspire creative human ideation for new design concepts. In the literature, the types and scopes of data used as design stimuli have varied. For instance, Liu et al. [19] proposed direct internet searches for web-based information to inspire product designers for concept generation. Goucher-Lambert & Cagan [20] showed that idea descriptions collected via online crowdsourcing campaigns can be used as design stimuli for expert designers. Purposefully curated design repositories, e.g., asknature.org and moreinspiration.com, which are structured databases to store and organize prior innovative designs or nature existences, have been proposed to drive concept generation [8,21]. One recent example is Idea-Inspire 4.0 [12], with a manually populated database of over 100 biological or engineered concepts, represented in the form of documents containing a function model, text, image, video, and audio together, as sources of inspiration for design concept generation.

Scholars have proposed the use of semantic networks to inspire designers for creative concept generation. For instance, Shi et al. [22] and Liu et al. [23] proposed the use of semantic networks mined from scientific papers as sources of inspiration for design concept generation. Chen & Krishnamurthy [24] proposed an interactive procedure to retrieve words and terms in ConceptNet to inspire designers. Doboli et al. [25] proposed a model to represent the great amount of knowledge in semantic networks in order to be used in creative activities, such as reasoning, decision making and design. Mukherjea et al. [26] extracted biological terms and their semantic associations from biomedical patent abstracts to build the BioMedical Patent Semantic Web to support knowledge discovery in the specialized biomedical domain. Song et al. [27] extracted the functional verbs from patent texts related to rolling robots and created a semantic network based on word co-occurrences in patents to identify core and peripheral functions in the network for recombination into new product ideas. Georgiev & Georgiev [28] has shown that semantic networks can be utilized to search for creative and successful ideas to support divergent thinking and generation of successful solutions to problems by using semantic similarity and information content based on WordNet's structure.

In contrast to open internet-based or manually curated design repositories, the patent database is a natural, and probably the largest, technology-based design repository, and big data source of design inspiration. It is organically accumulated and growing over time as inventors continually file patents for their inventions. Patent documents contain rich design information about the functions, components, structures and working principles of prior technologies and engineering designs and are organized in the international patent classification (IPC) system, which covers all domains of engineering and technology [29]. Prior studies on design ideation have demonstrated the efficacy of using patents as stimuli for the use of TRIZ or analogical designs [9,11,30–32]. However, prior studies have only drawn a small or arbitrary sample of patents as stimuli. The question of where in this vast and complex database to find and retrieve effective and novel stimuli for specific design interests or problems remains unanswered.

Recent studies based on human experiments have developed a nuanced understanding of human ideation behaviors and outcomes resulting from the use of different types of patents as inspirational design stimuli. Fu et al. [10] quantified the analogical distance between different patent documents based on latent semantic analysis and found that patents with shorter analogical distances to a given design problem can more effectively stimulate new ideas despite limited novelty, and patents analogically far from the design problem are ineffective in providing inspiration. In a different experimental setting, Srinivasan et al. [13] allowed engineering student designers to search for patents for inspiration by themselves to generate open-ended new ideas from a given original design object. They found that more proximate patent precedents are more likely to inspire designers and provide higher-quality inspiration, whereas precedents in far fields provide more novel inspiration.

The concepts of "near" and "far" stimuli were operationalized differently. Fu et al. [10] measured the analogical distance between patent documents based on the semantic similarity of the patent texts, whereas Srinivasan et al. [13] used an analogical distance between different design domains based on the similarity of patent references across patent classes. Despite the differences in experimental settings and distance metrics, the findings on the effects of using near-field versus far-field patents as design stimuli are aligned and resonate with the creativity literature, which has suggested that it is easier and more effective for humans to be inspired by external stimuli in fields near that of the original problem [10,13,33–36], while more distant stimuli may contribute to novelty with lower success rates [37–39]. In particular, the understanding of the multifaceted effects of the analogical distance on ideation behaviors and outcomes, based on the empirical contexts of patent data, may guide the retrieval of design stimuli from the vast patent database.

This literature review has suggested at least three research opportunities related to utilizing the entire patent database as a natural digital repository of design stimuli. First, the size and complexity of the patent database require strategies and guidance regarding which specific patent stimuli to retrieve given the interests of the designer and to provide inspiration with desired ideation outcomes. Prior studies utilized only a small sample of patent documents, instead of the total patent database, as the data sources, based on which the scope of inspiration might be limited. Second, while patent documents contain the rich details of engineering designs and were the design stimuli presented in most studies, patent documents are also difficult to read and comprehend and are thus unsuitable for rapid ideation. Alternatively, one can use keywords or terms extracted from patent documents as design stimuli directly. Reading terms for inspiration will be fast, but the inspiration can also be specific and unsystematic; there are millions of terms in the patent documents and retrieving them also requires a strategy. Third, a prior understanding of the effects of near- or far-field patents on creative ideation behaviors and outcomes may provide the theoretical grounding to build intelligent algorithms for the recommendation and guided retrieval of design stimuli from the patent database.

In this study, we aim to pursue these opportunities together in one knowledge-based expert system that provides guided exploration of semantic-, document- and field-level inspirational design stimuli from the total patent database. The guidance is based on the *knowledge distance* between technology fields.

**3. The Knowledge-Based Expert System**

3.1 Theoretical foundations

Our knowledge-based expert system is built on the fundamental understanding that new design ideas are conceived through the combination or analogical transfer of prior knowledge or concepts [33,40–45]. Combination refers to the cognitive process of bringing together multiple separate concepts to form a new one [42,46,47]. For example, the design concept of a

"flying car" is a combination of the airplane and car concepts. The design concept of a "smart watch" is a combination of the watch and smart phone concepts. Steve Jobs said, "creativity is just connecting things." Combination is a basic cognitive process of new idea generation and a natural feature of the associative memory of humans [40,48–50]. The effective combination of different concepts is enabled by their complementarity.

Analogy is the cognitive process of mapping the existing solutions to problems in a source domain to solve a similar problem in the target domain [34,37,51,52]. For example, the concept of a "bird nest" is mapped to the architecture domain to generate the design concept of the National Stadium of China. The "artificial neural network" concept in computer science is analogously drawn from the "neural network" concept in biology and brain science. Lidar, which was first invented to probe clouds and pollution in the atmosphere, was later analogously drawn to autonomous vehicle designs as a solution to probing the pedestrians on the streets. Design by analogy may be viewed as a specific type of combination between the solution in the source domain and the problem in the target domain. The effective analogical design is enabled by the similarity between the problems in the source and target domains.

Figure 1 depicts the human ideation process via the combination or analogy of prior knowledge or concepts across domains. However, for a creative designer or innovator, there is still the question of where the *source domains* are to find prior design concepts to combine with his/her current design for new ideas, or to find design stimuli related to solutions to the specific design problem he/she is faced with in the *target domain*. Many prior technologies have been invented in the past 2,000 years or more. Many, if not all, of these technologies and related knowledge are potentially useful for an engineer to combine into new ideas for his/her specific interests or to draw analogies in solving his/her specific problems. The key questions here are where to find prior knowledge to use as the design stimuli and how to retrieve it.

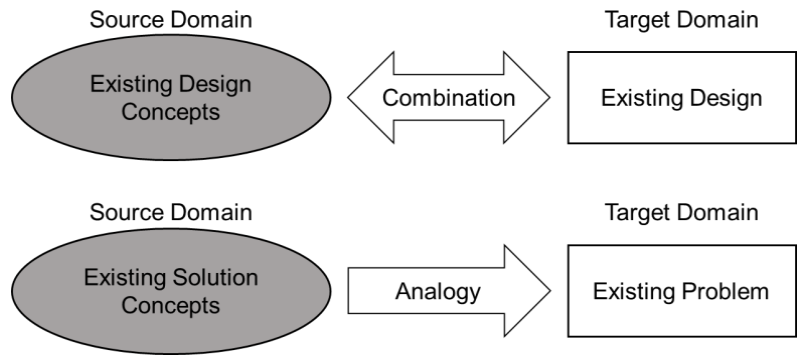

**Human ideation process via combination or analogy across domains**

Conceptually, all the technologies that humankind has created to date constitute the total technology space. The diverse knowledge about all the technologies forms the total technological knowledge space. The total technology space is a typical complex system [53] and constitutes many emergent fields and sub-fields of technologies. Likewise, the total technology knowledge space is also a complex system and constitutes many knowledge categories and sub-categories corresponding to different technology fields. This conceptualization is in line with category theory to represent the technological knowledge space as a complex system [54].

Such knowledge categorization or field specialization is naturally needed for effectively learning and reusing knowledge, given the scale and complexity of the total space and the cognitive and memory capacity constraints of individual humans and organizations. Given such human brain constraints, learning and design activities mostly take place within fields. Meanwhile, when we make inferences across different knowledge categories (corresponding to different technology fields) for learning, analogy, and combination, we may derive uncommon thinking and novel designs. And, such inferences across knowledge categories are facilitated by the similarity between the knowledge pieces in different categories [37].

In the technology knowledge space, two technologies or two fields of technologies are proximate if similar knowledge bases are required to design them and are distant if designing them requires distinct knowledge (as illustrated in Figure 2). Following learning theories

[37,55,56], it is easier and more effective for humans to understand and learn knowledge in the proximity of our own knowledge than knowledge that is farther away in the total knowledge space. Based on design creativity theories [10,13,16,35,37–39,57], for problem solving or innovation in a target domain, it is easier to obtain inspiration from nearer domains because knowledge base similarity facilitates inferences across domains. But stimuli from far domains lead to greater novelty and a greater chance of breakthrough innovation, despite the difficulty of making inferences across distant domains with dissimilar knowledge bases [37,39].

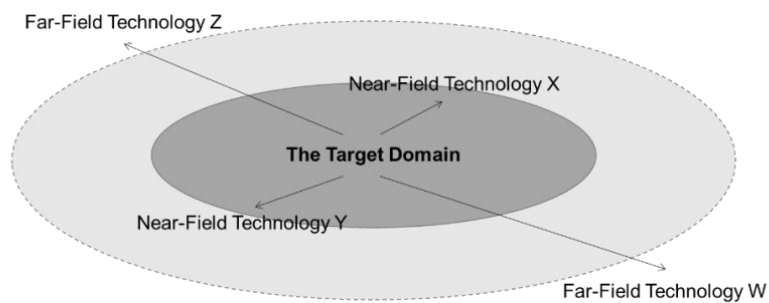

**Figure 2.** The technology space and knowledge distance between technologies

Therefore, we focus on *knowledge distance* (between the source domain of inspirational design stimuli and the target design domain) as the key variable to guide the search and retrieval of design stimuli for inferences across technology fields and combination- or analogy-based design ideation. Here, our concept of *knowledge distance* is broader than the analogical distance and combination distance concepts used in the prior literature and encapsulates them. Considering our interest in utilizing the patent database as the technology design repository (i.e., technological knowledge database), we operationalize the concepts of the total technology knowledge space and knowledge distance based on the patent database.

3.2 System design

Specifically, the *total technology space* is represented as a network of *all* patent classes (which are categories of patents) in the International Patent Classification (IPC), which cover all patent data and all fields of technologies. The patent classification system is a natural nested hierarchy to represent the structure of the total technology space and organize the data and knowledge about technologies belonging to different fields. At the top of the hierarchy, there are 8 categories, which are composed of 122 3-digit patent classes (e.g., aircraft, biochemistry and computing), which are themselves made of smaller classes with more nuanced 4- to 7-digit class codes. We have digitized the network map to enable zooming-in to show smaller technology fields defined by 4-digit patent classes and zooming-out to observe larger fields defined by 3-digit classes in an interactive manner similar to browsing Google Maps.

The *knowledge distance* is operationalized by the opposite concept of *knowledge proximity*, estimated as the Jaccard index (Equation 1), to associate technology fields represented by patent classes.

$$\varphi_{ij} = \frac{|C_i \cap C_j|}{|C_i \cup C_j|} \tag{1}$$

where $\varphi_{ij}$ denotes the knowledge proximity between fields *i* and *j*, $C_i$ and $C_j$ are the sets of patents cited by the patents in classes *i* and *j*, $|C_i \cap C_j|$ is the number of unique patents cited by the patents in both classes *i* and *j* (i.e., intersection), and $|C_i \cup C_j|$ is the number of unique patents cited by the patents in class *i* or *j* (i.e., union). The set of patents cited by the patents in a patent class approximates the knowledge base of the technology field represented by the patent class. Thus, Equation 2 quantifies the degree of overlap of the knowledge bases of two technology fields with a value between 0 and 1. While there exist alternative measures [58–61], this metric provides the greatest statistical explanatory power for inventors' historical exploration across technology fields based on patent data statistics [62–64].

To ensure statistical significance, the complete United States Patent & Trademark Office (USPTO) granted utility patent database from 1976 to 2018, with over 6 million patent documents and their references, is mined to estimate the knowledge distance between each pair of technology fields at the same level, defined in terms of the 3- or 4-digit IPC patent classes. Figure 3 shows the total technology space network of 3-digit IPC-defined technology fields positioned according to their pairwise knowledge distance. The size of a node corresponds to the total number of patents in the corresponding patent class since 1976.

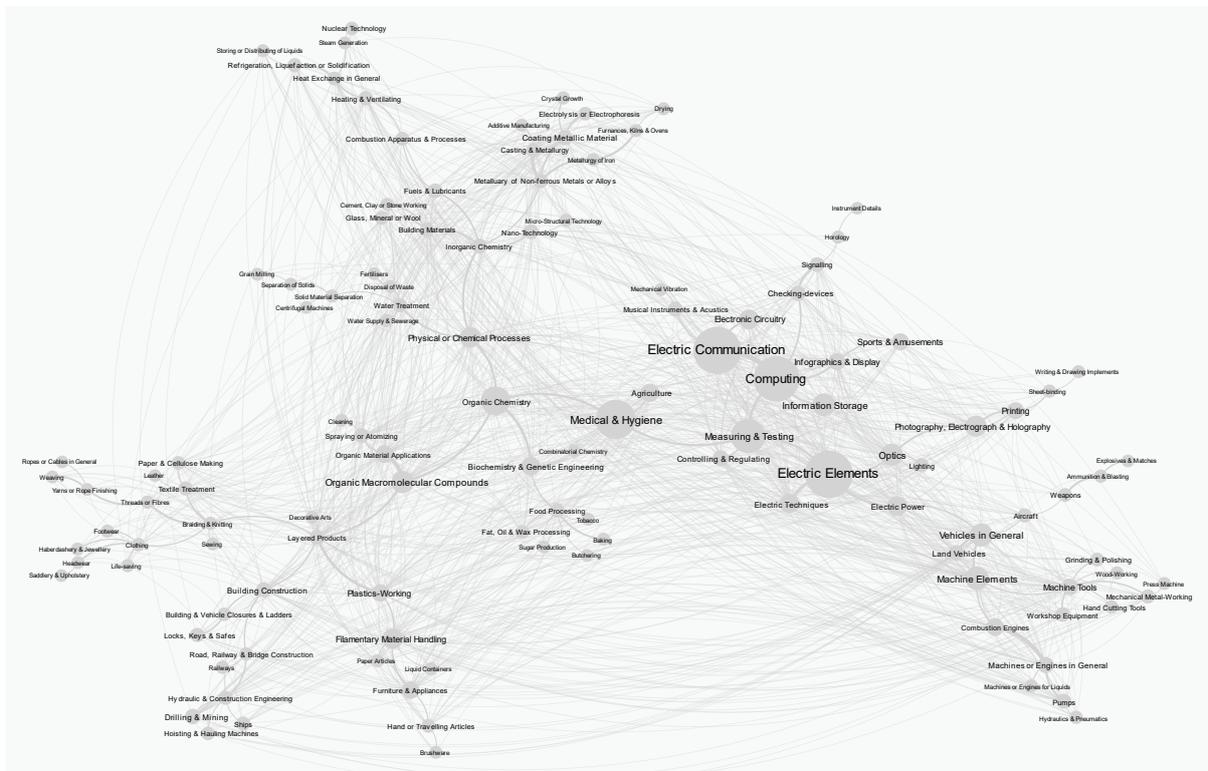

**Figure 3.** The total technology space map (visualized using a force-directed network layout).

The empirical network of technology fields is used as the structure of the relational database that stores prior technology knowledge data in different data categories corresponding to different technology fields and relates them according to the quantified knowledge distance. In turn, the network structure and knowledge distance information provide the basis for guiding inferences across fields to retrieve design stimuli stored in different fields (i.e., nodes) of the vast technology space (i.e., the network).

Specifically, to explore and retrieve design stimuli, one may first position the *target domain* of the original design problem or innovation interest in the total technology space, based on the classifications of the patents relevant to the problem or interest. The target domain may consist of one or multiple technology fields defined by IPC patent classes.[1] Thus, we quantitatively represent its position as a vector (***x***) of the distribution of the patents in respective patent classes, in the vector space (defined by the patent classes) that represents the technology space.[2] Visually, the target domain position can be shown by highlighting the network nodes that contain the relevant patents with a color intensity corresponding to patent counts.

---

[1] For example, the domain "flying car" spans across multiple IPC classes based on the classifications of the "flying car"-related patents, including "vehicles in general" and "aircrafts", in the network representation of the total technology space.

Once the target domain position is identified, in principle, design inspiration may be drawn from any field in the total technology space. To direct the exploration of potential design stimuli in near or far fields from the target domain (as illustrated in Figure 2), one strategy is to recommend the unexplored fields in "the white space" (i.e., the fields where no patent relevant to the original design problem or interest is found) according to their knowledge distance to the network position of the target domain, using the following formula:

$$\omega_j = \frac{\sum_{i \neq j} \varphi_{ij} x_i}{\sum_{i \neq j} x_i} \qquad (2)$$

where $\omega_j$ denotes the knowledge proximity of the unexplored white-space field *j* to the target domain; field *i* belongs to the set of fields in the target domain that contain patents directly relevant to the original design problem or interest; $x_i$ is the count of retrieved patents in field *i* that are relevant to the target domain; ***x*** is the position vector of the target domain in the total technology space; and $\varphi_{ij}$ is the knowledge proximity between technology fields *i* and *j*.

This metric quantifies the knowledge distance between a white-space field and the position of the target domain of the original design problem or interest in the total technology space. According to design creativity theories (see Section 3.1), the relatively proximate white-space fields are likely to provide the design stimuli that can be easily understood and inferred to the original design interest or problem for combination or analogy, while the prior design knowledge and concepts in more distant white-space fields may provide more novel inspiration despite a greater difficulty to perceive their relevance, analogy and combination.

---

[2] Similarly, the representation of the conceptual "source domain" may also cover multiple IPC-defined technology fields in the total network map.

Once a new design concept has been conceived with the inspiration of design stimuli from other fields, Equation 2 can be applied to estimate the knowledge distance between the inspiration source domain and the target domain of the original design problem or interest. This knowledge distance may inform the designer of the novelty of the new concept that he/she has just conceived in real time, based on design creativity theories. We had specifically tested the performance implications that can be drawn from the source-to-target knowledge distance information, which we quantified in our specific methodological framework.

In a human experiment with over 100 engineering students to generate innovative "spherical rolling robot" concepts [13], we asked the students to search for patent documents via Google Patents for design inspiration and to report the resulting concepts and the stimulating patents. More than 200 concepts were generated. Equation 2 was used to estimate the knowledge distance between the patent classes of the reported inspirational patents and the target domain ("spherical rolling robot"). We found patents in fields nearer to the target domain are more likely to inspire designers and stimulate ideas with higher quality, but patents in farther fields provide more novel inspiration. The results are aligned with design creativity theories but are based on using patents as design stimuli and the above knowledge distance quantification. They can provide actionable insights into what one can expect when using the total technology space map to explore design stimuli across different potential source fields.

From the feedback of the designers, we also learnt that patent documents are time-consuming to read, difficult to understand and can easily cause fixation. Therefore, to avoid fixation and provide rapid stimulation with diverse inspiration, our system retrieves design stimuli within each field at two granularity levels simultaneously: patent documents and terms representing the core design concepts of the field. The terms are extracted from the titles and abstracts of the patents classified in each field by using the TechNet (http://www.tech-net.org) and pre-stored in each node or category of the network database. TechNet has a very large vocabulary of over 4 million technical terms pre-extracted from all granted U.S. patents from 1976. It outperforms WordNet, ConceptNet and other semantic databases in technical term retrieval and inference tasks [29]. The terms in TechNet represent functions, components, structures and working principles in engineering designs and provide specific design information. Thus, they allow designers to quickly obtain highly specific inspiration from a large quantity of diverse technical concepts.

To facilitate the search and retrieval of patent documents and technical terms from different technology fields, we utilize the cloud-based InnoGPS system. The interactive visualization features of InnoGPS were designed by analogy to those of Google Maps. While Google Maps is used to position buildings (geographical objects), exploring neighborhoods, and determining directions to far locations in the physical space, InnoGPS allows for positioning technologies (and related companies and persons), exploring neighborhoods and determining directions to far locations in the technology space. An early version of InnoGPS was used to position companies on a map according to their patent records for diversification analysis [65] and to analyze the degrees of competition among different firms [66].

Here, we supplement InnoGPS with the functionality of storing and retrieving the design concept terms and patent documents in different patent classes to stimulate creative design ideation of inferences, analogies or combinations across technology fields based on knowledge distance. We call this expanded knowledge-based expert system for design ideation "InnoGPS 2.0". Compared to the prior version, InnoGPS 2.0 is aimed to stimulate more granular ideation in individual persons to generate more nuanced design concepts. Figure 4 depicts the architecture and core elements of the system, which enables the following computer-aided ideation workflow:

1) position the target domain (of design problem or interest) in the total technology space,
2) explore the technology space to identify the inspiration source domains,

3) exploit specific design stimuli (design concept terms or documents) within a source domain for inferences, analogies and combinations with the original design problem or interest.

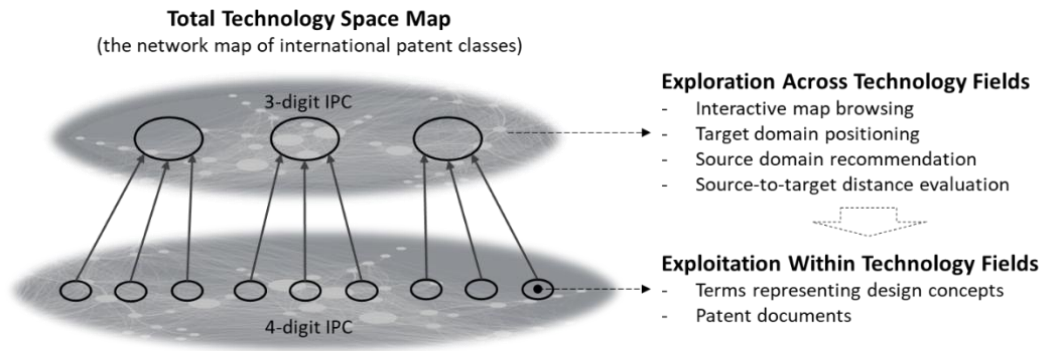

**Figure 4.** Architecture and core elements of the proposed knowledge-based expert system

**4. Case Studies**

Now we present two case studies using the knowledge-based expert system to retrieve design stimuli to aid human ideation on 1) open-ended product innovation ideas around rolling toys and 2) solution ideas to a specific design problem, namely, water seepage in subway tunnels.

4.1 Computer-aided ideation for new, open-ended design ideas

The first case is aimed at answering the innovator's question "*What new design concepts can be combined with our prior product*?" Assume that we were designers and product managers working in a small "rolling toy" company (e.g., https://www.sphero.com/). We previously had a successful "*rolling toy*" product, which reached its market limit. Its success attracted many competitors to offer similar products. We need to design and launch new products beyond but around the existing "rolling toy" designs. To meet such a need, traditionally, one could use expert panels, brainstorming, user studies, market research and so on to search for technologies for potential combination with the "rolling toys". Such processes are slow, time consuming and require manpower and capital resources. Here, we use our knowledge-based expert system to search for design stimuli and aid in open-ended "rolling toy"-related design idea generation in a rapid manner.

The first step is to "position" the prior technologies related to "rolling toy" in the network (Figure 5), using the upper-left search bar in the interface. The positioning function searches for the set of patents in the total patent database with the keyword "rolling toy" in their titles and abstracts, identifies the classifications of retrieved patents, and determines the position vector of "rolling toy" in the total technology space based on the distribution of "rolling toy"-related patents in different patent classes. In this case, "Sports & Amusement", "Infographic & Display", and "Vehicle in General" are highlighted in red because "rolling toy" patents are found in these 3-digit IPC-defined fields, and they constitute the target domain for potential design inferences. The intensity of the red color of the nodes corresponds to the rolling toy patent occurrences in the respective patent classes. The gray nodes contain no "rolling toy"-related patents and constitute the "white space" for the "rolling toy" design.

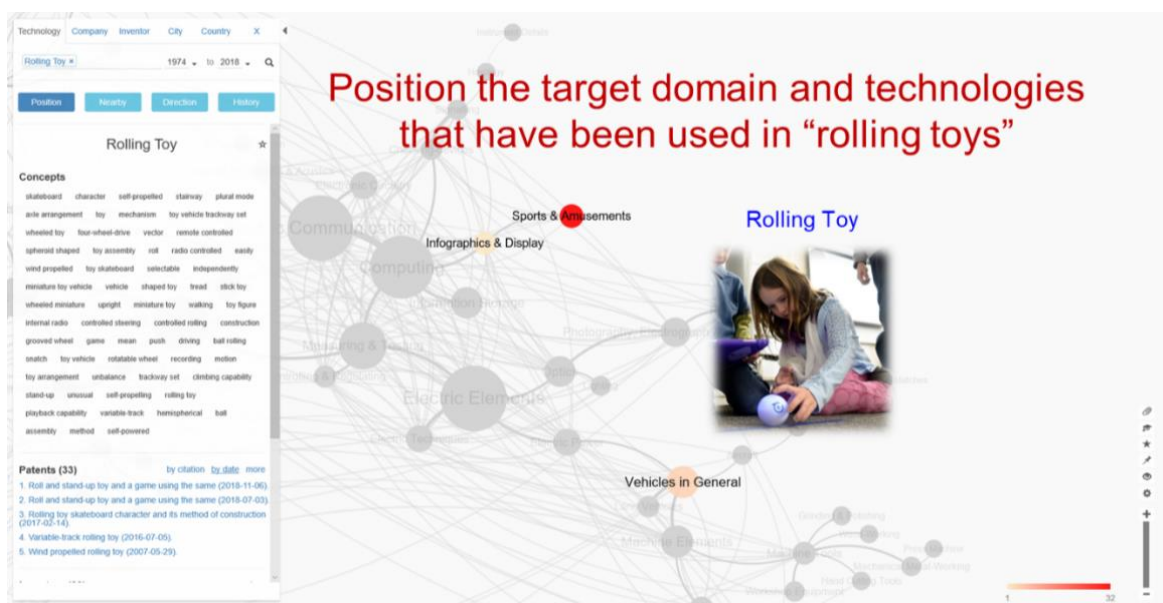

**Figure 5.** Positioning technologies related to prior rolling toy designs

The information panel on the left-hand side lists the technical terms that represent the core design concepts used in prior rolling toy designs and the patent documents that represent the prior inventions related to rolling toys, as well as the related inventors and companies. Such information describes the concepts, designs and technologies that have been adopted in rolling toy designs. The terms can be sorted by their occurrence frequencies in the domain, indicating how representative the terms are of the domain. The patent documents can be sorted according to grant dates or cumulative citations. The most recent patents may represent the innovation frontier of rolling toys, and the most cited patents may represent the foundational technologies for rolling toy designs. With such information, we can make more informed decisions regarding differentiating our own designs from these prior works to innovate and avoid competition.

The knowledge-based expert system is most useful for the exploration of the white space, i.e., the gray fields in the space, for innovation. The "nearby" function, based on Equation 2, quantitatively identifies, and visually highlights the unexplored fields in the white space that are most proximate to the target domain, i.e., the red nodes. As shown in Figure 6, one can move the slider bar in the information panel to increase the knowledge distance of the exploration into the white space from the nearest to the farthest fields relative to the target domain, "Rolling Toy". The nearest white-space fields may contain novel but easy-to-learn design stimuli for potential combination with rolling toys to generate new product ideas. In this case, "Checking Devices" appears to be the nearest white-space field to the target domain.[3]

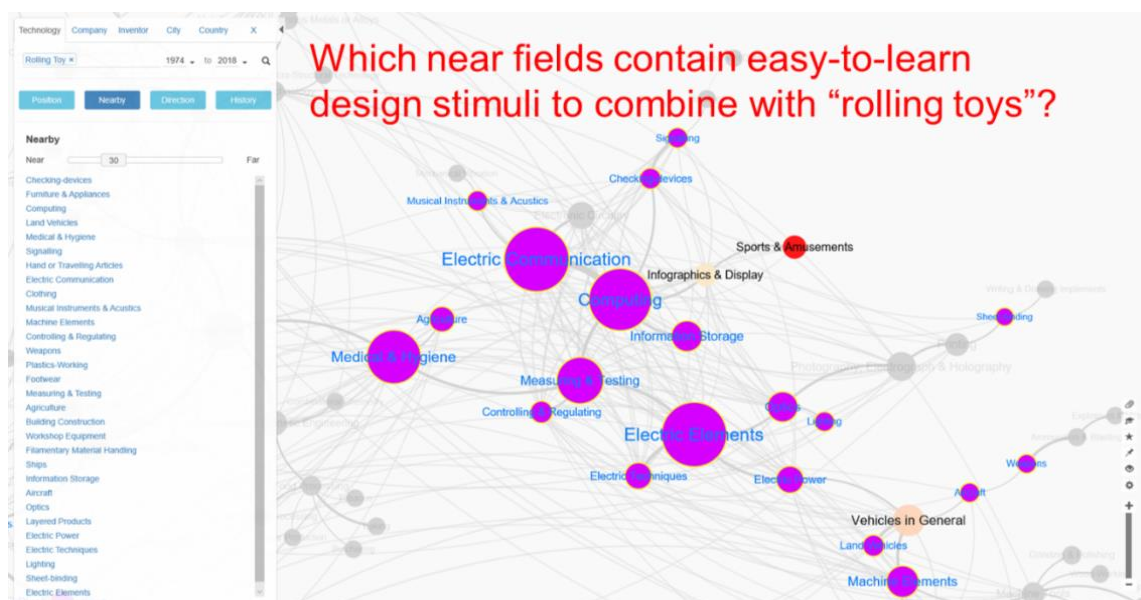

**Figure 6.** White-space fields near the target rolling toy design domain

---

[3] Based on Equation 2, the knowledge proximity between "Checking Devices" and "Rolling Toy" is 0.081366.

We click the node "Checking Devices" on the map to activate the information panel for this specific field, defined by a 3-digit IPC class. The panel reports the leading inventors and companies as well as the most cited and newest patents in the field. Most importantly, the panel reports the terms that occur most frequently in the patent documents and represent the core design concepts (e.g., characteristic functions, components, structures and working mechanisms of technologies) in this specific field (Figure 7).[4] Instead of reading patent documents for design inspiration, we quickly go through these elemental design concepts and combine them with the rolling toy to generate new product ideas.

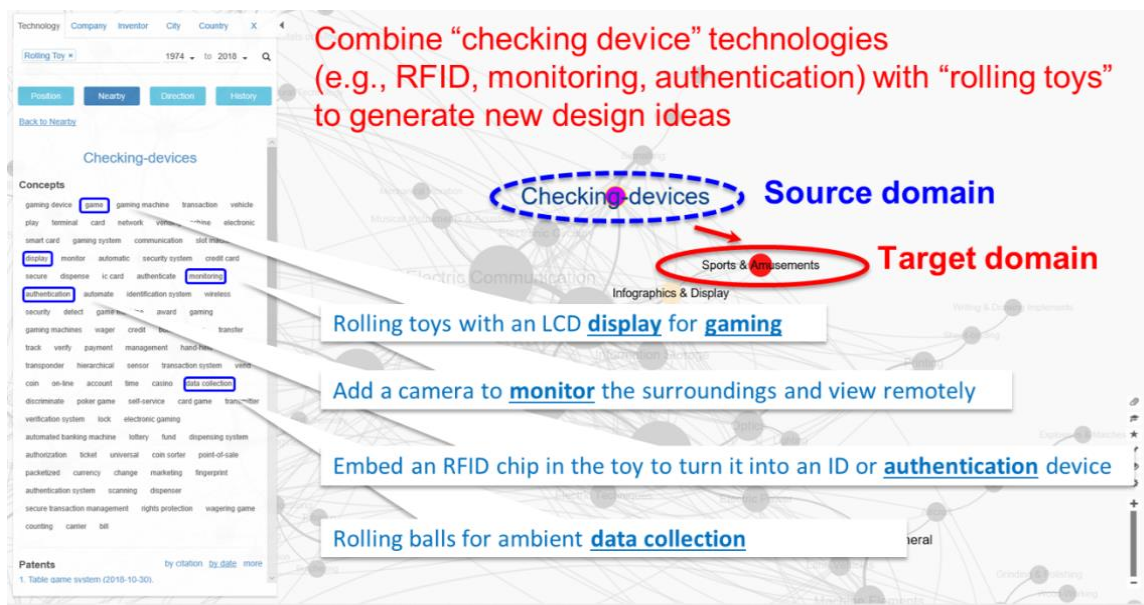

**Figure 7.** Combine concepts in the checking-device domain with rolling toy designs

For instance, in the panel, the terms "gaming" and "display" inspired us to develop a new design of rolling toys combined with an LCD display for children to play interactive visual games on the rolling ball. The concept "monitor" inspired the combination of a camera with the rolling toy to monitor surroundings remotely. The "authentication" function can be combined with a rolling toy by embedding a radio frequency identification (RFID) chip inside it so that it can be used as an authentication device. Children could use playful rolling balls as their identity keys in kindergarten. Furthermore, the terms "sensors" and "data collection" inspired us to conceive of rolling balls combined with relevant sensors that cruise and collect environmental data in difficult-to-visit places, such as underground sewers.

---

[4] The terms can be rank ordered by term frequency that indicates the popularity of terms in prior designs of this field, or by weighted term importance indexes such as TFIDF. Alternatively, the terms can be rank ordered by their semantic similarity to the key terms in the target domain.

In addition to the most characteristic terms, one can also directly read the patent titles (which are single short sentences) for rapid inspiration. For example, the patent title "dual-mode vehicular controller" stimulated us to conceive of the design concept of "dual-mode" rolling toys that have a mobile-controlled mode and an autonomous mode for rolling. The patent title "tamper resistant rugged keypad" inspired us to conceive of the design concept of rolling toys with "rugged" shells, while the surfaces of existing rolling toys are normally smooth. Without reading the lengthy full texts and images for patents, we are able to obtain inspiration to rapidly generate various divergent design ideas that we would not be able to conceive of without the prompts of the terms or sentence-level stimuli from the knowledge-based expert system. These new design concepts are highly specific and can be readily prototyped because of the granularity of the stimuli.

In addition to the nearby fields, one can freely browse the map regardless of the knowledge distance to discover and seek inspiration. For instance, noticing the field node "Mechanical Vibration"[5] on the map, we conceive of the abstract idea that some technologies in the "Mechanical Vibration" field might be potentially combined with rolling toys to generate entertaining vibrations as a new function. "Lighting" is another 3-digit IPC-defined field.[6] When seeing the "lighting" node on the map, we were inspired and generated the idea of combining lighting technologies with rolling toys to provide mobile lighting at home or in public spaces. In these cases, the stimulation is at the field level, e.g., lighting, or mechanical vibration, and the generated ideas are about the macro-level design direction and are not specific enough for prototyping or implementation.

---

[5] Based on Equation 2, the knowledge proximity between "Mechanical Vibration" and "Rolling Toy" is 0.003978.
[6] Based on Equation 2, the knowledge proximity between "Lighting" and "Rolling Toy" is 0.029773.

Once the designer conceives of a high-level idea with stimuli at the field level, through either heuristic map browsing (e.g., the "Lighting" and "Mechanical Vibration" examples) or system recommendations based on the knowledge proximity to the target domain (e.g., the "Checking Device" example), he/she can further click the field node on the map to activate the information panel of the corresponding inspirational field in order to discover generic design concepts for more specific and nuanced design stimulation. For instance, when we entered the "Lighting" field to retrieve the characteristic design concepts there, the prompted term "LED" stimulated a more specific idea of incorporating LEDs into a spherical rolling ball to provide moving lighting (Figure 8). This new idea is finer grained and can be readily prototyped. These examples reveal the differences between ideas generated with term- and field-level stimulation.

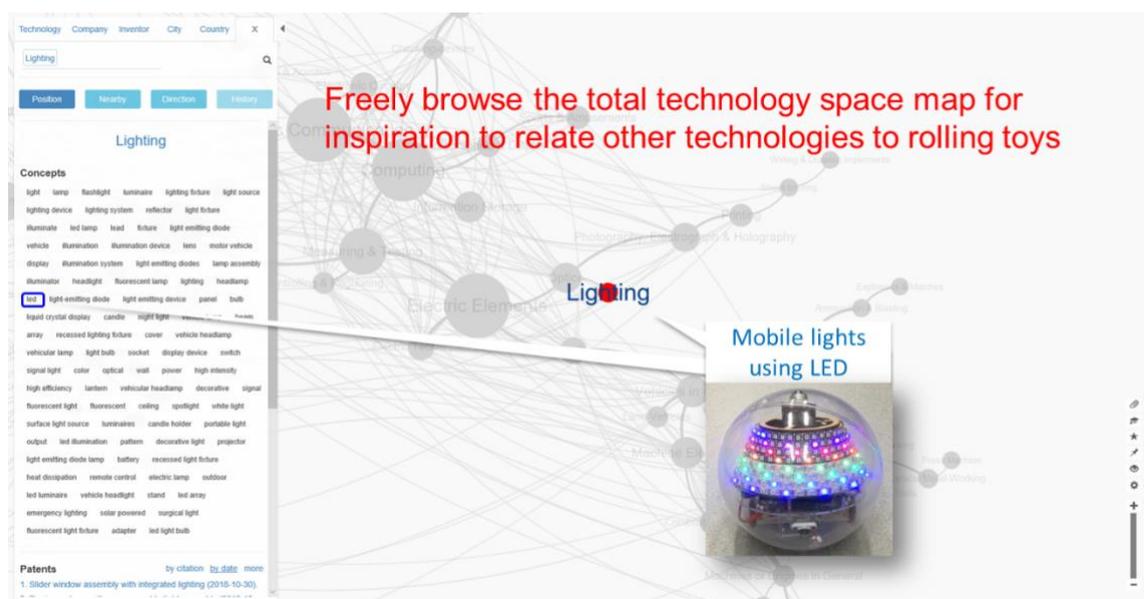

**Figure 8.** Free map navigation for design inspiration for rolling toy designs. The prototype image is from a student design team in the SUTD Engineering Design Innovation course.

While heuristic map browsing may stimulate one to generate diverse design concepts with inspiration from different fields in the total technology space, the knowledge distance from the inspiration source domain to the target design domain (e.g., rolling toy in this case) indicates both the novelty of a new idea in the patent space and the feasibility of realizing the idea. As suggested by prior design creativity studies [10,13,35], new ideas that combine near-field concepts might be more feasible, whereas far-field stimuli contribute to novelty. With the understanding of such tradeoffs, as shown in Figure 9, the map-based visual information regarding how far or near a source domain of design stimuli is from the target domain may allow for the instant evaluation and comparison of different design ideas generated in the map-aided rapid ideation process.

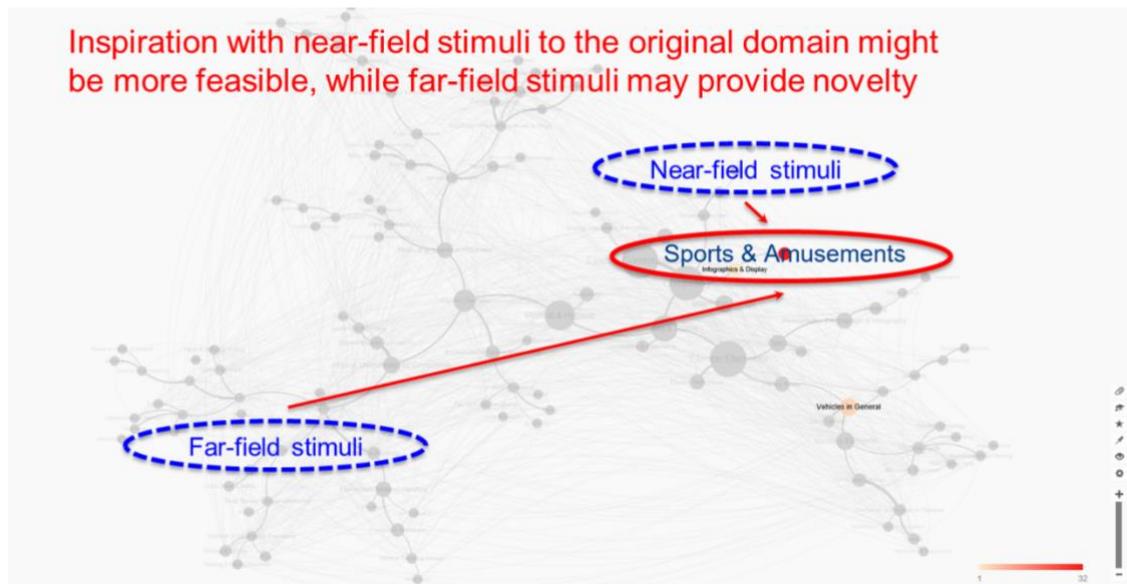

**Figure 9.** Exemplar design concepts with stimuli at different distances to the original design

It is particularly noteworthy that the novelty and patentability of new ideas are naturally ensured when they are generated by combining design stimuli from a white-space field. Additionally, the expert system also provides rapid quantitative evaluation of the relative novelty of different new design ideas based on the quantified knowledge distance. For instance, one can store the generated ideas in their respective inspiration fields on the map (i.e., in the corresponding data categories in the network database). Figure 10 is the interface for reporting and sorting the new ideas generated for rolling toy innovation, according to the knowledge proximities between their inspiration source fields and the target domain. Table 1 reports the knowledge proximity values based on Equation 2.

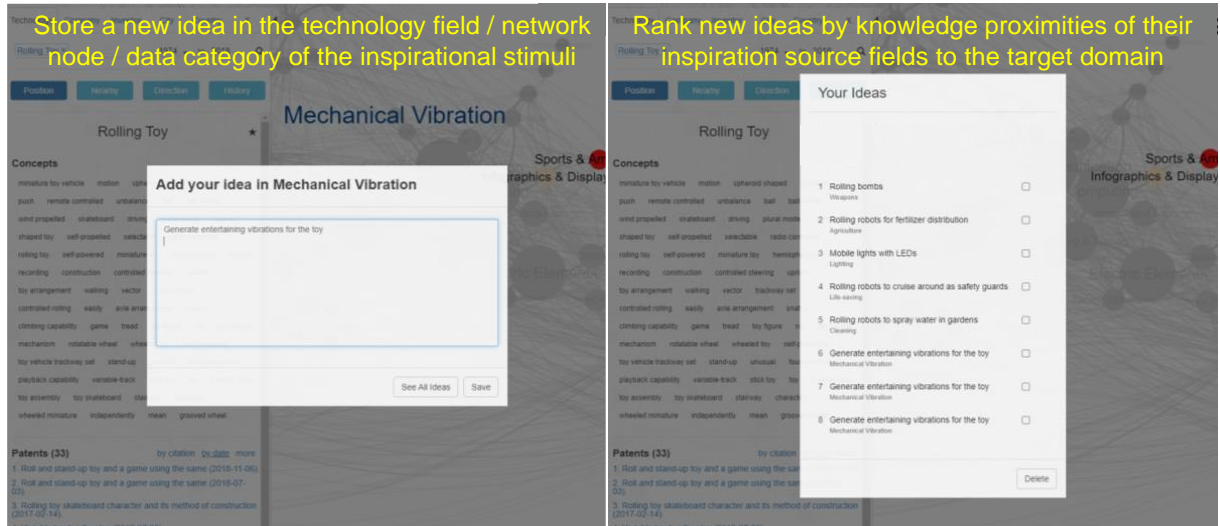

**Figure 10.** New idea storage in knowledge category of stimuli and ranking by knowledge distance

**Table 1.** Knowledge proximities of the inspiration fields of new ideas to the target domain

| Inspiration Field ($j$) | Target | New Ideas | Proximity ($\omega_j$) |
|---|---|---|---|
| Weapons | Rolling Toy | Rolling bombs | 0.044023 |
| Agriculture | Rolling Toy | Rolling robots for fertilizer distribution | 0.041288 |
| Lighting | Rolling Toy | Mobile lights with LEDs | 0.029773 |
| Life-saving | Rolling Toy | Rolling robots to cruise around as safety guards | 0.020480 |
| Cleaning | Rolling Toy | Rolling robots to spray water in gardens | 0.007609 |
| Mechanical vibration | Rolling Toy | Generate entertaining vibrations for the toy | 0.003978 |

4.2 Computer-aided ideation for design problem solutions

The second case study answers a different innovator's question: "*Existing subway tunnels face water seepage over time; what can we design to solve the problem*?" This was a real problem statement provided by the Land Transport Authority (LTA) of Singapore to the faculty and researchers at Singapore University of Technology and Design (SUTD) for design solution ideas. When a company or organization is faced with such a problem, the traditional response has been to consult experts, conduct brainstorming among peers, crowdsource ideas from the public and so on to explore potential solutions. Such approaches rely on social processes and are time consuming and serendipitous. Here, we use our knowledge-based expert system to aid in the generation of design ideas to solve the given problem.

Again, the first step is to position the technology fields in which solutions to the general "water seepage" problem have been previously developed. Figure 11 shows that relevant technologies are found in 23 fields (i.e., 3-digit patent classes) across the total technology space. They are highlighted in red color of varied intensity indicating the frequency of "water seepage"–related patents in respective patent classes. Hereafter we refer to this set of technology fields as the "red space". In the red space, the "Building Construction", "Hydraulic and Construction Engineering", and "Road, Railway and Bridge Construction" fields appear the reddest, as they contain the largest numbers of "water seepage"–related patents. The gray domains have no prior patents related to water seepage.

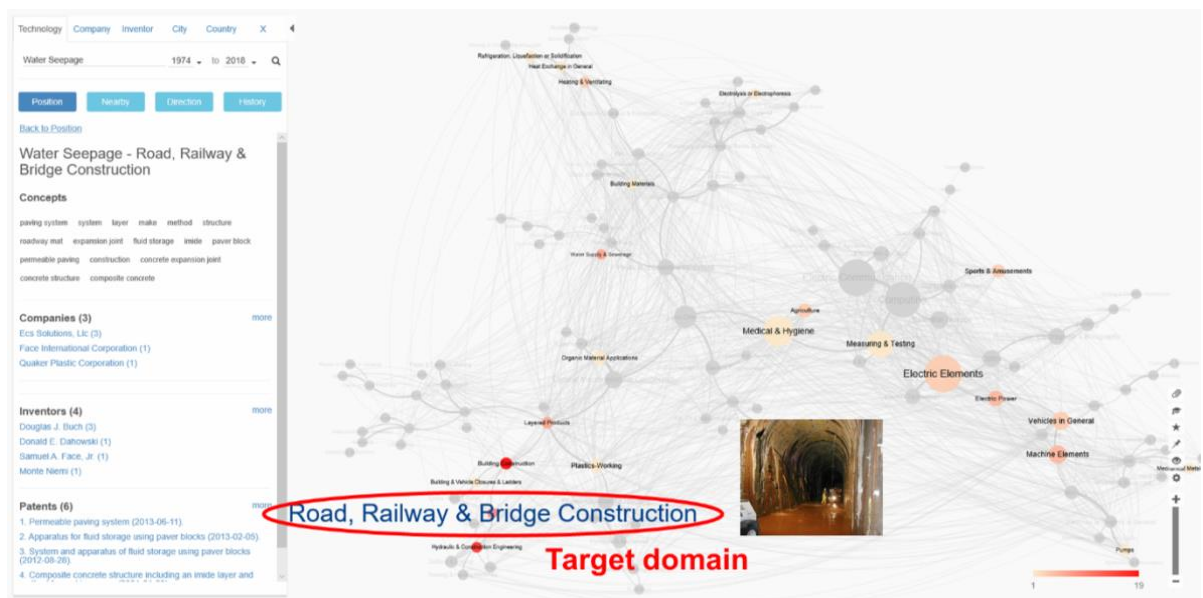

**Figure 11.** Identifying the "red space" domains with solutions to "water seepage"

The information panel reports the terms extracted from the "water seepage"–related patent titles, sorted by their occurrences in the patent set, and the "water seepage"–related patent titles with hyperlinks to the full documents, sorted by either citation counts or recency, as well as a list of inventors and organizations according to their relevant patent counts. These items inform us of the concepts and technologies that have been adopted in "water seepage"–related designs. Furthermore, the panel also provides a list of technology fields that contain "water seepage"–related designs, corresponding to the fields highlighted in red on the map.

Among the fields that contain "water seepage"–related designs, "Road, Railway & Bridge Construction" (at the bottom left corner of the total technology space map) is the field of specialization of the LTA of Singapore and is where the design problem is situated. This field is considered the *target domain* in our theoretical framing in Section 3.1. With this understanding, we can explore existing solutions to the general "water seepage" problem in other red-space fields as potential source domains and leverage them to solve the "water seepage" problem in the designer's target domain. If a "water seepage" solution in another field has not been used in the target domain, the analogy (or the solution-problem combination) will be novel. Such design-by-analogy opportunities exist because different solutions to the same high-level problem have been created by specialized designers in different fields.

The "water seepage"–related design concepts in the nearest red-space fields to "Road, Railway and Bridge Construction" can be most easily learned and mapped to the target domain. For instance, we can click the field node "Layered Products"[7] on the map to activate the information panel for this red-space field defined by a 3-digit IPC class. In this case, the panel retrieves and reports only the "water seepage"–related design concepts and patent documents (as well as their inventors and the organizations that own them) within the "Layered Products" field. By contrast, if we click a white-space field node (as demonstrated in the rolling toy case), the information panel retrieves and reports all the design concepts and patent documents within that white-space field as potential design stimuli.

In the "Layered Products" field, only four patents related to "water seepage" were found and retrieved. We opened each of the patent documents to read the design details for inspiration related to preventing water seepage. For instance, the patent entitled "*water barrier panel*" (granted in 1977) describes a panel of two sheets filled with a composition containing bentonite, a water-soluble dispersant, and a water-soluble polymer (Figure 12). The drawings in the patent document are very helpful for understanding the design. From the patent document, we learnt the bentonite composition that is seepage resistant. With this inspiration, the resulting design idea is to dispose a mass of swellable bentonites to the areas of possible water seepage in the wall of the subway tunnel. In this design by analogy, the source domain is "Layered Products", and the target domain is "Road, Railway and Bridge Construction". A water seepage prevention solution is mapped from the source domain to the target domain.

---

[7] Based on Equation 1, the knowledge proximity between "Layered Products" and the target domain is 0.026634.

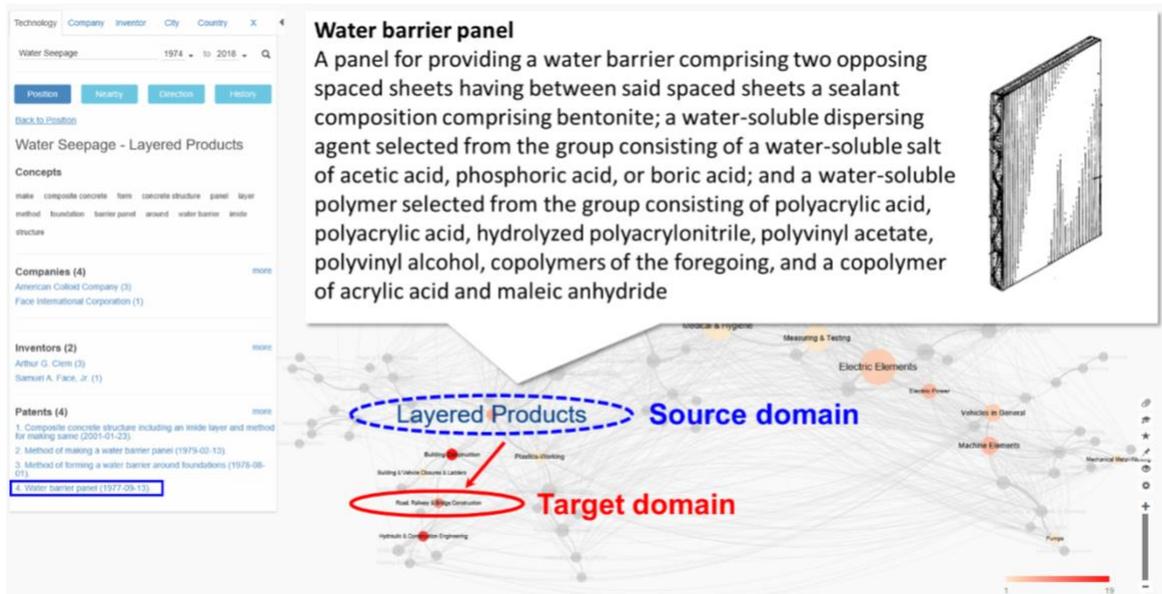

**Figure 12**. Patent document as inspirational stimuli for design by analogy

When we click to exploit the "Organic Material Applications" field,[8] which is more distant from the target domain ("Road, Railway and Bridge Construction") than the "Layered Products" field, only one "water seepage"–related patent is found and retrieved (as shown in Figure 13). The patent document is entitled "*sealing ponds and reservoirs by natural materials*" (granted in 1987) and describes a method to produce a water-impervious layer with a mixture of natural materials (sand, montmorillonite, and salt water) to decrease hydraulic conductivity, for use at the bottom of ponds and water reservoirs to prevent water seepage into soil structures. This patent document contains no drawing but specific and detailed composition of the mixtures to produce the water-impervious layers. By analogy, this specific water-impervious layer made of natural materials can be applied to the walls of subway tunnels to prevent water seepage from the soil structures to the tunnel.

---

[8] Based on Equation 1, the knowledge proximity between "Organic Material Applications" and the target domain is 0.017903.

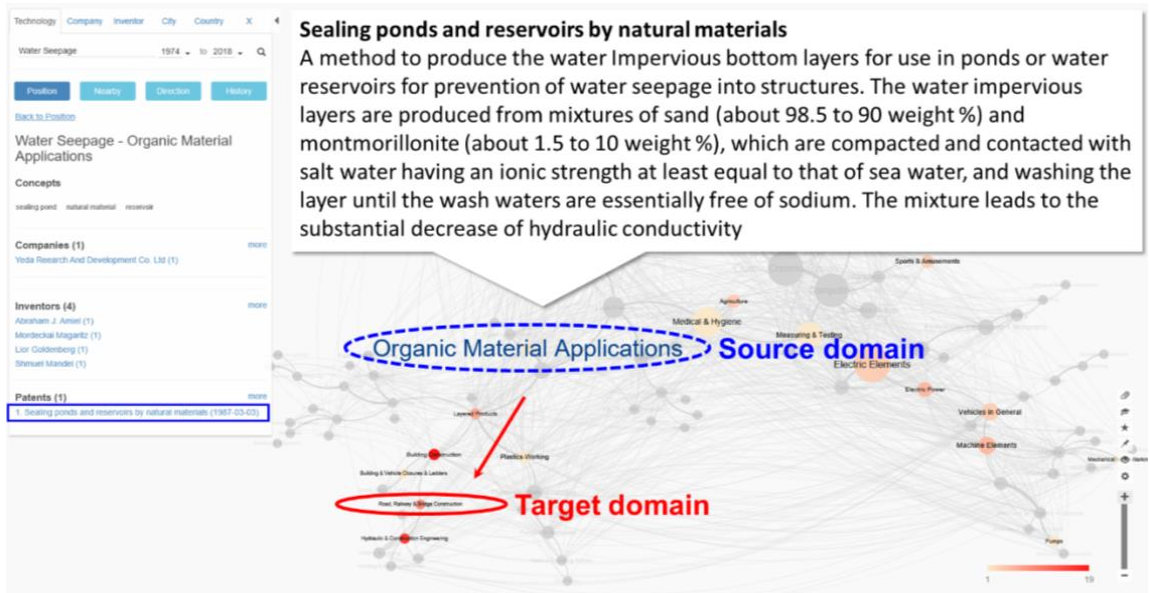

**Figure 13**. Patent document as inspirational stimuli for design by analogy

In an even more distant field to the target domain, "Medical & Hygiene",[9] only one "water seepage"–related patent is found and retrieved (as shown in Figure 14). It is entitled "*appendage covering system*" (granted in 2013) and describes a structure to protect the appendage of a user from the contamination or water seepage while bathing or outdoors during inclement weather. The design includes a protective cover and a sleeve made of an impermeable plastic (as shown in the patent drawings, Figure 14). With this inspiration, we generate a new idea of placing a circular impermeable plastic structure to cover water leaking spots in the ceiling of the subway tunnel and direct the water to the side drainage.

In this case study, we have focused on demonstrating the retrieval and use of patent documents as design stimuli. Meanwhile, the technical terms are also retrieved and reported simultaneously with the patent documents in the process (see left-hand side information panels in the figures above). One can also seek inspiration from the list of technical terms representing the elemental concepts used in prior solutions to the same "water seepage" problem in different technology fields. While drawing inspiration from the terms can be faster, patent documents provide more detailed and nuanced inspiration, as shown in the computer-aided ideation process above. Therefore, the knowledge-based expert system provides design stimuli at different granularity levels that complement each other.

---

[9] Based on Equation 1, the knowledge proximity between "Medical & Hygiene" and the target domain is 0.008192.

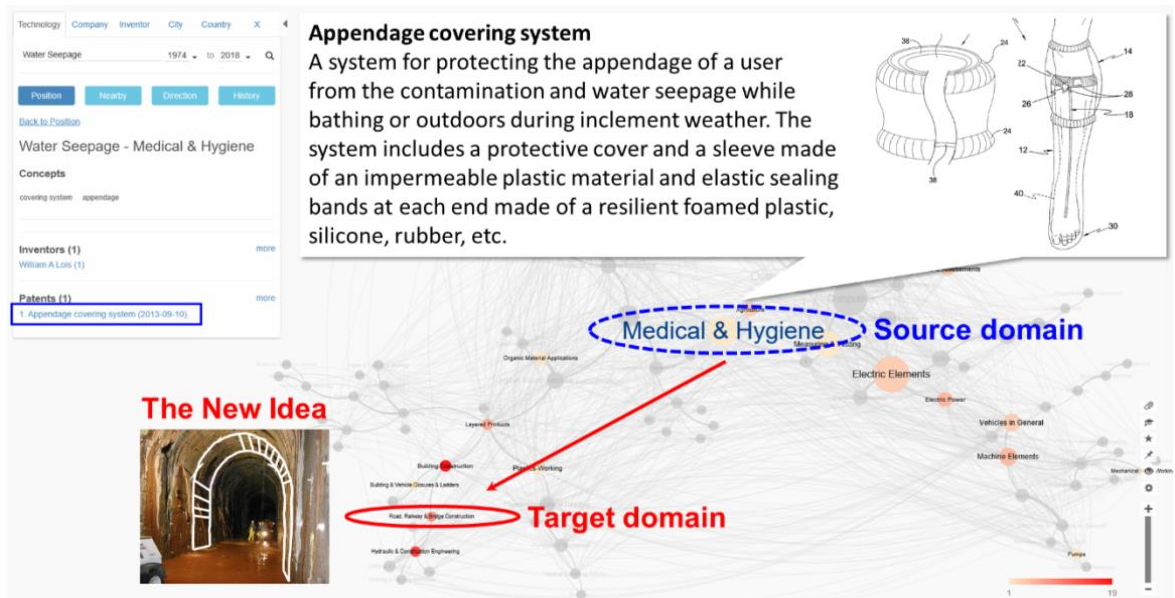

**Figure 14**. Patent document as inspirational stimuli for design by analogy

## 5. Discussion

The two case studies show the effective uses of the proposed knowledge-based expert system to explore and exploit different forms of design stimuli (i.e., field, document and term), support different cognitive ideation processes (i.e., combination and analogy), and answer different design innovation questions (i.e., open-ended innovation and specific problem solving). Both case studies demonstrate the same general process – starting with positioning the target domain in the total technology space, then exploring the total space to identify inspiration source domains, and finally exploiting and using specific design stimuli within source domains to generate new design ideas.

In the rolling toy case, we generate new, open-ended ideas around a given prior design. But we generate solution ideas to a given problem in the water seepage case. We explore the white space for design stimuli in the rolling toy case, whereas we explore the red space in the water seepage case. In the rolling toy case, new ideas are mainly generated via the combination of retrieved prior design concepts (e.g., "data collection" function or the "LED" component) in the white-space source fields and "rolling toy", whereas in the second case, new ideas are generated by drawing analogies and leveraging solutions to similar problems in the source domains to solve a given problem in the target domain. In the first case, we demonstrate the use of terms as design stimuli, while patent documents are used as stimuli in the second case.

Moreover, in the first case, one could also explore the red space, seek inspiration from patent documents and generate new "rolling toy"–related ideas by analogy, and in the second case, one could also explore the white space, seek inspiration from terms and generate new ideas via basic combinations. The variations in the two cases were aimed at covering broad scenarios of a computer-aided design ideation process, aided by our knowledge-based expert system. In both cases, the design heuristics (combinations and analogies across domains) are supported and empowered by the knowledge-based expert system, which guides the designer to retrieve design stimuli (across the term, document and field levels) from the total patent database and seek inspiration across fields according to the knowledge distance.

Generally, when the design stimuli and inspiration are drawn from a more distant source domain to the target domain, the resulting design ideas obtained via combination or analogy are expected to present greater novelty but poorer quality and feasibility, as suggested by the prior design creativity literature. This understanding can inform the search for inspiration across near and far fields regarding potential outcomes when one uses the expert system's visual and quantitative information about the relative distances or proximity between the source domains of the design stimuli and the target domain of the original design problem or interest during the computer-aided ideation process.

Overall, this work has made at least three contributions to the studies of data-driven design and knowledge-based expert systems for design. First, in prior studies, the provision of design stimuli was at either the concept, document, or field level, but not all together. Concept terms may provide specific inspiration rapidly but lack details [23,67]. Patent documents may provide rich design details for systems, products, and processes but requires more time to read and efforts for comprehension that may cause fixation [13,68]. A map of technology fields may avoid ideation fixation by broadening the search and provide rapid inspiration for design directions, but the resulting ideas are macro-level and not specific enough for implementation [66,69]. Our system synthesizes the provisions of rapid design stimulation from concept terms, nuanced and systematic design stimulation from patent documents, and macro-level inspiration for design directions from the field nodes in the total technology space map during the computer-aided ideation process.

The second contribution lies in the use of the total patent database as the source of design stimuli. This choice is driven by the spirit of comprehensiveness. In contrast, prior patent studies focused on the retrieval and analysis of a sample of patents relevant to a specialized topic or domain, providing limited ranges of design stimuli. The digital patent database covers all fields of technology and is growing every day, and in principle, it may stimulate design ideas by analogy or the combination of knowledge or concepts across far or near fields to varied degrees in the total technology space.

Meanwhile, the total patent database is vast and demands scientifically grounded rules and systematic guidance to enable the retrieval of potential design stimuli from it. To address this challenge, we use a technology space network constructed on the international patent classification system as a digital infrastructure to store, organize and guide the retrieval of design stimuli data in different patent categories according to the quantified knowledge distance. The focus on the knowledge distance to organize the categories of design data allows the retrieval process to be grounded in design creativity theories and the previously established understanding of the effects of using far or near patent stimuli for design ideation. This constitutes our third contribution.

For instance, the visual and quantitative information of the distances between different fields in the network may guide the designer to explore stimuli in fields either near or far relative to his/her home design domain with the awareness of the potential ideation performance outcome. Also, it may quickly inform the designer of the relative novelty and feasibility of different new design ideas when they are generated with stimuli from source fields in the white or red space with different knowledge distances to the target domain. Such instant and theoretically grounded guidance may ease later idea evaluation and selection efforts. In contrast, traditional ideation techniques, when used by inexperienced designers without extensive knowledge, are more likely to generate ideas that have already existed elsewhere or are highly infeasible because the idea conception process does not ensure novelty, provide indications of the potential ideation outcomes of different choices of stimuli, or enable the instant evaluation of conceived new ideas. Such noisy ideas will need to be evaluated and screened during a later concept selection process, requiring additional workloads. In this sense, our system and process supports the "design for patentability" strategy for inventions [70].

Furthermore, our work may also inspire transfer learning research in the field of deep learning [71–75]. Instead of transferring trained learning models from a source domain to a target domain for classification tasks (i.e., transfer to learn), our system guides transferring general knowledge concepts across domains to create new design concepts (i.e., transfer to create). Similar knowledge-based systems to ours, with an empirically derived network architecture of knowledge categories or domains, may be developed to organize the source domains with trained learning models and guide the transfers of them to target domains according to inter-domain knowledge distance. Such a knowledge-based system may also provide knowledge distance-based guidance to "transferring to create" using generative models as CycleGAN [76] that map and fuse data across domains, in terms of which source domains to choose to transfer and the potential novelty and fidelity of the generated designs.

## 6. Limitations and Future Work

A few limitations of the present system exist and suggest directions for future research. First, there exist alternative knowledge distance metrics. Future work may allow the automatic selection of an optimal distance metric for a specific case context based on the historical data of the case on demand. Second, the current system only implements one network visualization layout for either 3-digit and 4-digit IPC classes. Future research should experiment different visual layouts and allow mapping with finer-grained 5-digit to 7-digit classes to enable the exploitation of small, specialized areas in the total technology space. Third, images related to specific concepts may help designers understand the meanings of such concepts (as illustrated in our second case study) and thus can be matched together with the prompted concept terms to enhance stimulation. Research to retrieve patent images as design stimuli is on-going [77].

The current paper has primarily focused on demonstrating the procedures and effectiveness of using the proposed system to aid in idea generation. We do so via two detailed case studies. Moving forward, we plan to conduct controlled experiments with large groups of designers with different backgrounds and statistically measure their performances in terms of the quantity, novelty, quality, speed, and breadth of the generated ideas, as well as reactions of the designers. In particular, there are other emergent processes of new idea generation [51,78] than combination and analogy focused on cross-domain knowledge transfers [79], e.g., design heuristics [2], TRIZ [9], other data-driven methods [12,19,49,80] and serendipity [46,51]. Thus, we plan to compare the performances of experimental groups using different approaches and processes to generate ideas. Meanwhile, we also plan to conduct more advanced use cases that move downstream in the design innovation process to produce actual designs, porotypes and inventions for users and experts to test, evaluate, and choose, beyond the ideas.

Other than patent data, design stimuli can also be drawn from proprietary enterprise data or public data sources, such as technical papers, reports, web articles, books, course documents, and product descriptions on e-commerce sites. In fact, these diverse unstructured non-patent data about technologies can be also classified and stored into respective IPC-defined categories, which are associated and organized in the empirically identified, latent but natural technology network structure. Specifically, neural networks can be trained on patent texts (and images) and their classifications (as labels) and then used to classify non-patent data. That is, the technology space network may serve a general digital infrastructure to store and organize the world's data of technological knowledge and concepts according to their fields, and to guide the retrieval of them according to the quantified knowledge distance among fields.

Furthermore, the current system architecture also allows several artificial intelligence capabilities to be added to the system, such as 1) machine learning algorithms to learn and understand a user's latent design preferences and aptitudes from his/her digital footprints, 2) intelligent recommendations of fields, documents, and concepts according to user preferences and aptitudes, and 3) automatic idea generation algorithms. For instance, natural language generation algorithms can be developed to create sentence descriptions of new design ideas like those we generated in the case studies. This can be achieved by using syntactic structure templates for certain design heuristics (e.g., combination or analogy) to synthesize the computer-retrieved stimulating terms (e.g., "data collection") and the designer's starting design object or problem (e.g., "rolling toy"). Table 2 presents two examples of automatic idea description generation.

**Table 2.** Idea generation using idea description templates

|  | Example 1 | Example 2 |
|---|---|---|
| Design heuristics | Combination | Analogy |
| Template | "Combine [*a design concept in the source domain*] with [*the design object in the target domain*]" | "Adopt [*an existing solution in the source domain*] to solve [*a given problem in the target domain*]" |
| Original design topic or problem | "*Rolling toy*" | "*Water seepage in subway tunnels*" |
| Design stimuli | "*Data collection*" | "*Composite concrete layer*" |
| Idea description | "Combine *data collection* with *rolling toy*" | "Adopt *a composite concrete layer* to solve *water seepage in subway tunnels*" |

Such capabilities enable *computer ideation* beyond *computer-aided ideation of humans*. Clearly, automatic computer ideation, being tireless and faster than human ideation, has natural advantages for the wide exploration of many macro fields and the deep exploitation of many micro and specific design stimuli throughout the total technology space. Then, human ideation can focus more on the interpretation, evaluation, selection and adaptation of computer-generated ideas rather than the search and identification of stimuli. These constitute our longer-term research plan of developing expert systems with *creative artificial intelligence* beyond computer-aided human ideation.

## 7. Conclusions

In this paper, we have proposed an knowledge-based expert system to provide computationally guided exploration and exploitation of design stimuli taken from the total patent database at the concept, document and field levels simultaneously, according to the inter-field knowledge distance, to support analogy and combination-based creative design ideation. We also demonstrated the computer-aided ideation processes, with the proposed knowledge-based expert system. The system aims to enhance design ideation and concept generation at the fuzzy front end of the innovation process and make it more informed, inspired, and rapid by incorporating big data and intelligent guidance based on creativity theories. The knowledge-based expert system architecture is scalable and allows the inclusion of broader non-patent natural-language data sources as well as machine learning and artificial intelligence capabilities for computer ideation beyond computer-aided ideation in the future. Therefore, we hope this research is not viewed as a conclusion but as an invitation for further research and development of knowledge-based expert systems that augment design creativity and enhance innovation.